\documentclass[%
  aps,
  prl,
  reprint,          
  superscriptaddress,
  floatfix,
  amsmath,amssymb,
  nofootinbib
]{revtex4-2}

\usepackage[T1]{fontenc}
\usepackage[utf8]{inputenc}
\usepackage{lineno}

\usepackage{graphicx}
\usepackage{siunitx}
\usepackage{bm}
\usepackage{physics}    

\begin{document}

\title{A geometric basis for materials families in inorganic solids}

\author{Justin Tahmassebpur}
\altaffiliation{*Corresponding author: jt577@cornell.edu}
\affiliation{Department of Applied and Engineering Physics, Cornell University, Ithaca, NY, USA}
\author{Sarvesh Chaudhari}
\affiliation{Department of Physics, Cornell University, Ithaca, NY, USA}
\author{Crist\'obal M\'endez}
\affiliation{Department of Applied and Engineering Physics, Cornell University, Ithaca, NY, USA}
\author{Rushil Choudhary}
\affiliation{Department of Physics, Cornell University, Ithaca, NY, USA}
\author{Sudipta Kundu}
\affiliation{Department of Chemistry, Pennsylvania State University, State College, PA, USA}
\author{Raymond E. Schaak}
\affiliation{Department of Chemistry, Pennsylvania State University, State College, PA, USA}
\author{H\'ector Abru\~na}
\affiliation{Department of Chemistry and Chemical Biology, Cornell University, Ithaca, NY, USA}
\author{Peter Frazier}
\affiliation{School of Operations Research and Information Engineering, Cornell University, Ithaca, NY, USA}
\author{Tom\'as Arias}
\affiliation{Department of Physics, Cornell University, Ithaca, NY, USA}

\date{\today}

\begin{abstract}
The thermodynamic stability of inorganic solids spans a vast compositional space, yet materials scientists have long organized their intuition around a manageable number of materials families. Here we show that this organization has a precise geometric basis. The formation-energy convex hull of all inorganic compounds from the Materials Project, spanning 92-dimensional elemental composition space, is captured to near DFT accuracy by a polyhedron with only seven facets. Each facet corresponds to a family of materials sharing similar chemical potentials. This low-dimensional structure is not merely an economical description of energies: without retraining or structural input, the same framework reproduces trends in DFT-calculated defect energies and elemental spatial correlations in high-entropy nanoparticles. These results reveal that a small number of material families, corresponding to geometric features of composition–energy space, govern bulk stability, defect energetics, and elemental mixing, and provide a unified, interpretable framework for rapid screening across diverse materials systems.
\end{abstract}

\maketitle

\renewcommand{\thefootnote}{\fnsymbol{footnote}}
\footnotetext[1]{Justin Tahmassebpur (corresponding author): jt577@cornell.edu}

Inorganic solid-state materials encompasses an enormous diversity of compounds, yet materials scientists navigate this complexity with remarkable efficiency~\cite{curtarolo2013highthroughput,materials_project}. Concepts such as ionic salts, transition-metal chalcogenides, Zintl phases, and intermetallics organize vast swaths of materials space into coherent families with shared bonding character, stability trends, and functional properties. This organizational power is so deeply embedded in practice that it can seem self-evident---a matter of pedagogy rather than physics. Here we show that it is, in fact, a quantitative reflection of the geometry of the thermodynamic stability landscape itself: the formation-energy convex hull of inorganic solids~\cite{materials_project,saal2013materials} is far simpler than its combinatorial definition would suggest, and that simplicity is precisely what makes materials families not just useful, but inevitable.

Formally, the convex hull may be viewed as a function of composition, $\mathrm{Hull}[E(\boldsymbol{x})]$, where $\boldsymbol{x}$ denotes the stoichiometry of a material. Here, $\boldsymbol{x}$ is a 92-dimensional vector of elemental fractions, $\mathrm{H}_{x_1}\mathrm{He}_{x_2}\ldots\mathrm{U}_{x_{92}}$, and the formation energy is defined as
\begin{equation}
E(\boldsymbol{x}) \equiv \min_n E(S_n(\boldsymbol{x})),
\end{equation}
the minimum over all locally stable atomic arrangements $S_n(\boldsymbol{x})$ at fixed composition. In principle, identifying this minimum requires exploring a combinatorially large set of metastable structures for each composition~\cite{oganov2006crystal,woodley2008crystal}. 
Large-scale density functional theory (DFT) efforts, including the Materials Project and others~\cite{materials_project,saal2013materials,chanussot2021oc20,tran2023oc22,Curtarolo2012AFLOWLIB,Choudhary2020JARVIS}, have made remarkable progress in sampling this space and constructing extensive databases of formation energies. Bonding-geometry aware models, including graph neural networks (GNNs), trained on these databases now achieve near-DFT accuracy for formation energies and related properties across diverse materials classes~\cite{xie2018crystal,schutt2018schnet,chen2019graph,gasteiger_dimenet_2020,cheng_crystal_2022,Choudhary2021ALIGNN}. Generative models, such as diffusion-based crystal generators~\cite{mattergen}, further accelerate discovery by proposing plausible candidate structures. However, both first-principles calculations and their learned surrogates remain tied to the underlying search over ${S_n(\boldsymbol{x})}$: even when candidate structures are generated efficiently, many must still be evaluated to identify the lowest-energy phase at fixed composition. From this perspective, the combinatorial complexity of atomic structure space appears unavoidable, suggesting that the convex hull should be a highly complex function of composition.


Against this expectation, an unexpected result has emerged: formation energies can often be predicted accurately from composition alone, without any explicit structural information. Beginning with ElemNet~\cite{jha2018elemnet} and followed by Roost~\cite{goodall2020predicting}, IRNet~\cite{sun2020irnet}, and CrabNet~\cite{Wang2021crabnet}, deep neural networks have demonstrated that composition-only inputs suffice to reproduce DFT-calculated formation energies across vast materials spaces. These models are trained to approximate the full composition--energy function $E(\boldsymbol{x})$, from which the convex hull is derived and which ultimately governs thermodynamic stability. Their success is striking: they achieve high accuracy despite never exploring the exponentially large set of atomic configurations $\{S_n(\boldsymbol{x})\}$ that formally defines $E(\boldsymbol{x})$. Yet the reason for this success---what structure in the formation-energy landscape makes composition alone sufficient---has remained unexplained. This observation suggests that the mapping from composition to formation energy, and in particular the convex hull $\mathrm{Hull}[E(\boldsymbol{x})]$, may possess an intrinsic simplicity not evident from its combinatorial definition. This raises a foundational question: \emph{is the convex hull of formation energies governed by a simple, interpretable function of composition?}

Here, we show that the answer is \emph{yes}. We show that the Materials Project formation-energy convex hull, spanning more than 50{,}000 bulk compounds, can be remarkably well approximated by a polyhedron with only \emph{seven} facets, yielding a low-dimensional geometric description of materials stability across broad chemical space. We further show that this seven-facet thermodynamic structure extends beyond bulk stability itself. Without retraining or structural input, the same composition-only framework reproduces trends in DFT-calculated defect energetics and experimentally observed elemental spatial correlations in high-entropy nanoparticles, indicating that the same small set of thermodynamic families organizing formation energies also organizes defect energetics and complex-material mixing behavior. Together, these results reveal an unexpected simplicity in the organization of inorganic materials thermodynamics and suggest a route to rapid, interpretable screening in settings governed by energy differences, such as the design of oxygen diffusion barrier materials~\cite{chaudhari2025active}.

\section*{Results}
\begin{figure*}[htbp]
  \includegraphics[width=0.99\textwidth]{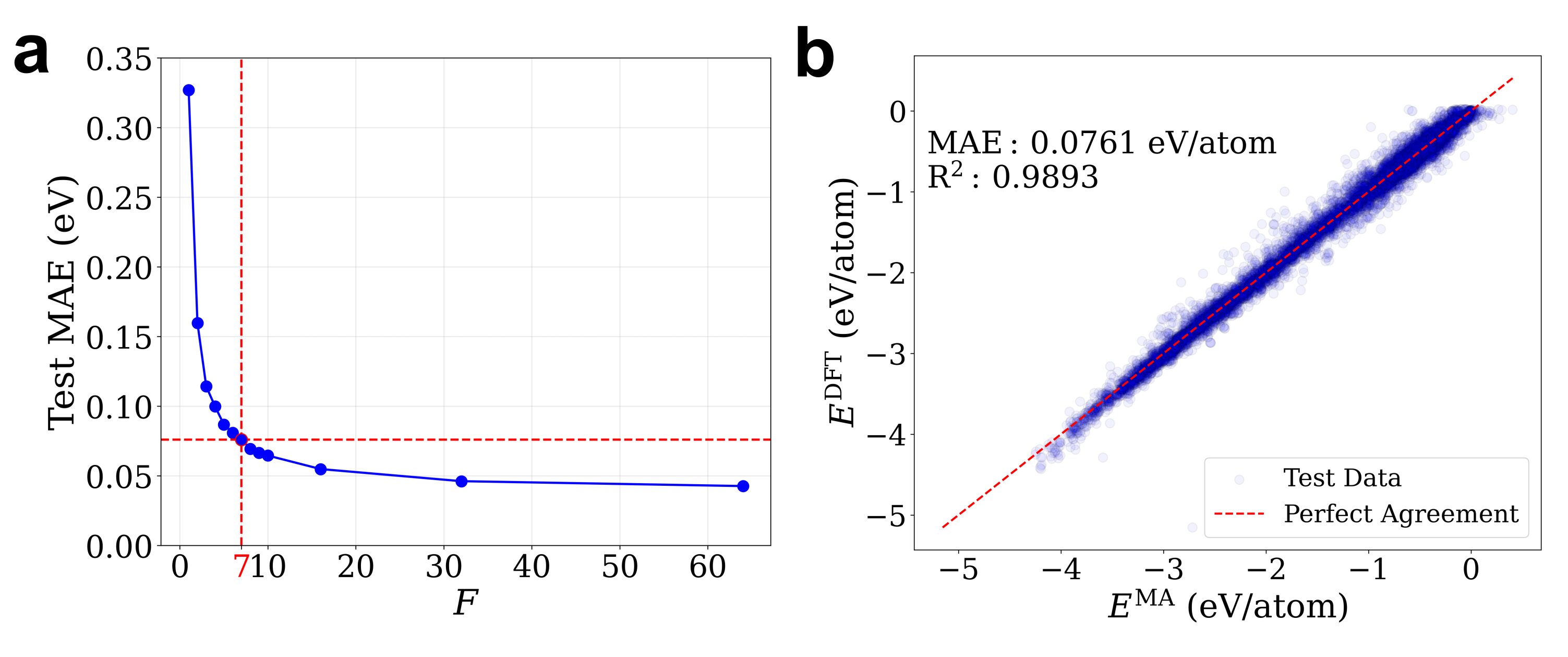}
  \caption{\textbf{Performance of max-affine representation for prediction of formation energies.}
Mean absolute error (MAE) on the 20\% test set for the max–affine representation as a function of the number of facets $F$, with the chosen value $F=7$ indicated by the red lines.
\textbf{b} Max–affine representation values for $F=7$ on the 20\% test set versus DFT-computed formation energies from the Materials Project~\cite{materials_project}. The red line indicates perfect agreement ($y=x$).}
  \label{fig: formation}
\end{figure*}

\subsection*{Low-complexity structure of the formation-energy convex hull}

By construction, the convex hull $\mathrm{Hull}[E(\boldsymbol{x})]$ is a convex function of composition. Convex functions admit a natural geometric description in terms of supporting hyperplanes---geometrically, planes that touch the surface from below without crossing it---and max–affine representations provide a corresponding piecewise-linear parameterization in which the function is expressed as the maximum over a set of affine (i.e., planar) components. In this picture, the convex hull is represented as a polyhedral surface composed of such supporting facets. For a formation-energy hull spanning such a high-dimensional and composition space spanning diverse materials classes as that of all inorganic solids, one would expect this surface to require a very large number of facets to be represented accurately.

To probe this expected complexity, we represent the formation-energy convex hull using a max--affine function
\begin{equation}
    E(\boldsymbol{x}) = \max_{f=1}^{F} \left( \sum_{s=1}^{92} a_{f,s} x_s \right) + b,
    \label{eq: max-affine}
\end{equation}
where the input $\boldsymbol{x}$ is the 92-dimensional composition vector of elemental fractions, $a_{f,s}$ and $b$ are fitted coefficients, and $F$ is the total number of supporting facets. We fit this representation using the lowest-energy Materials Project structure for each composition lying within $0.025$~eV/atom of the convex hull, giving 51{,}068 compounds. The data are split randomly into an 80\% subset used to determine the parameters and a 20\% subset used to evaluate the accuracy of the representation. Figure~\ref{fig: formation}a shows that the approximation error for the test set decreases rapidly with the number of facets and is already close to saturation at just $F=7$. The corresponding seven-facet representation (Fig.~\ref{fig: formation}b) achieves a mean absolute approximation error of $0.076$~eV/atom for the test data not used in the fit, approaching DFT accuracy. Although a larger number of facets would be required to reproduce the exact hull everywhere, these results show that the dominant near-hull thermodynamic structure is governed by only seven supporting hyperplanes.

The error decay in Fig.~\ref{fig: formation}a is approximately a power-law, with fitted exponent $\alpha \approx 1.2$ for an error of the form $c + bF^{-\alpha}$. This convergence is dramatically faster than the worst-case max--affine approximation rate predicted by theory~\cite{balazs2015maxaffine}, which in our 92-dimensional composition space would scale as roughly $F^{-2/92} \approx F^{-0.021}$. The observed scaling therefore deviates significantly from generic high-dimensional behavior, indicating that the formation-energy convex hull is far simpler than a typical convex function in 92 dimensions and instead possesses a genuinely low-complexity global structure.

\begin{figure*}[htbp]
  \includegraphics[width=0.99\textwidth]{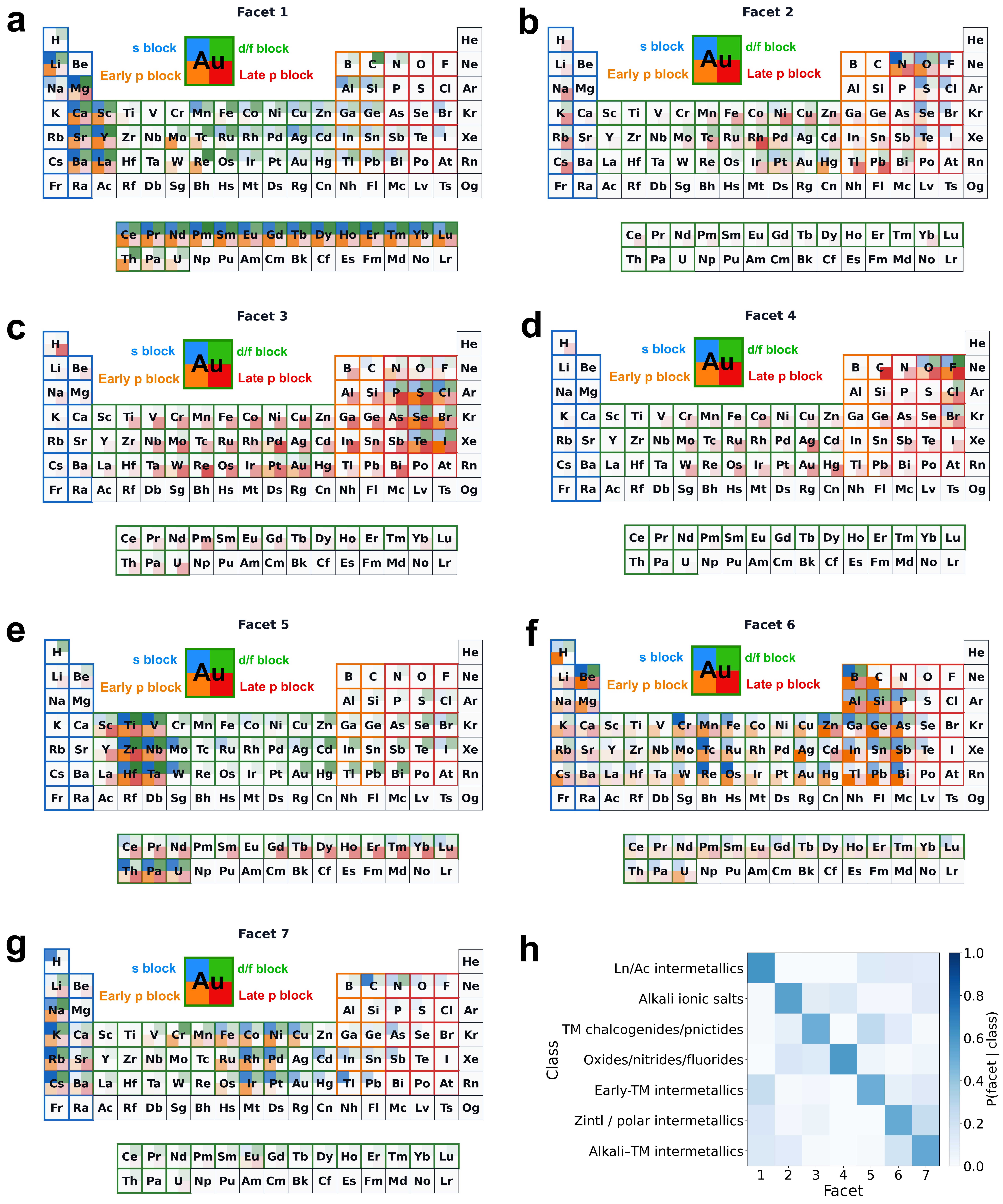}
    \caption{\textbf{Interpretation of max-affine facets for binary compounds.}
    \textbf{a--g} Facet-resolved periodic-table maps for facets~1–7 of the max-affine hull. As indicated by the enlarged element key, the border color of each element denotes its valence category (blue: $s$ block; green: $d$/$f$ block; orange: early $p$ block; red: late $p$ block). Within each element box, the four colored quadrants indicate how frequently that element forms binary compounds with elements from each category on the corresponding facet. 
    \textbf{h} Facet–materials-class correlation matrix, where color intensity denotes the probability that a binary compound from a given materials class activates each facet.}
  \label{fig: interpretation}
\end{figure*}

\subsection*{Materials families from convex hull geometry}
The emergence of materials families in inorganic solids points to an underlying regularity in the organization of thermodynamic stability across composition space. We now show that this regularity is directly reflected in the structure of the convex hull and admits a simple physical interpretation within the max–affine framework.

In the max--affine representation, each material composition is associated with a single active facet of the hull. For a material with composition $\mathbf{x}$, the active facet is the facet that attains the maximum value at that composition,
\[
f^\star = \mathrm{argmax}_f\, \sum_{s=1}^{92} a_{f,s} x_s.
\]
Within that facet, the associated coefficients define effective compositional chemical potentials
\[
\tilde{\mu}_s = \frac{\partial E(\boldsymbol{x})}{\partial x_s} = a_{f^\star,s},
\]
where $E$ is the formation energy per atom. All materials with compositions lying on the same facet therefore share the same facet-specific effective chemical potentials. Each facet therefore defines a distinct thermodynamic regime with a characteristic thermodynamic fingerprint.

If these facet-defined regimes correspond to genuine thermodynamic structure, they should align with recognizable materials families. To test this, Fig.~\ref{fig: interpretation} analyzes all stable and nearly stable binary materials within 0.025~eV/atom of the Materials Project convex hull. We group elements into four valence categories,
\[
C\in\{s,\ d/f,\ \mathrm{early}\ p\mathrm{\ block},\ \mathrm{late}\ p\mathrm{\ block}\},
\]
and, for each facet $f$, quantify how frequently each element $E$ forms binaries with elements from category $C$. Specifically, we define $w_{E,C}^f$ as the sum of the atomic fractions of $E$ over all binaries on facet $f$ in which the other element belongs to category $C$. We then normalize across facets,
\[
\tilde{w}_{E,C}^f = \frac{w_{E,C}^f}{\sum_{f'} w_{E,C}^{f'}},
\]
and finally rescale within each facet by
\[
I_{E,C}^f = \frac{\tilde{w}_{E,C}^f}{\max_{E',C'} \tilde{w}_{E',C'}^f}.
\]
The quantity $I_{E,C}^f$ therefore measures how strongly element $E$ is represented on facet $f$ in binaries with elements from category $C$.

Figs.~\ref{fig: interpretation}a--g render these quantities as periodic tables for facets 1--7. The border color of each element indicates its own valence category, while the four interior squares indicate the values of $I_{E,C}^f$ for the four possible partner categories. In this way, each panel summarizes the characteristic bonding environments associated with a given facet.

The resulting maps show that the seven facets align closely with recognizable materials classes, rather than forming arbitrary partitions of composition space. Facet~1 is enriched in lanthanide and actinide intermetallics with strongly metallic $d/f$--$d/f$ and $d/f$--$s$ bonding. Facet~2 highlights alkali ionic salts, such as Li--O/N/S/Se/F/Cl/Br compounds. Facet~3 captures transition-metal chalcogenides and pnictides, while Facet~4 corresponds primarily to mixed ionic--covalent oxides, nitrides, and fluorides involving $p$-block and transition-metal cations. Facet~5 is dominated by early transition-metal intermetallics. Facet~6 is enriched in binaries between alkali or alkaline-earth metals and early $p$-block elements, including many Zintl-like materials, polar $s$-$p$ intermetallics, and covalent semiconductors. Facet~7 reflects metallic alkali--transition-metal and alkali--alkali binaries.

To quantify this correspondence between facet-defined regimes and materials families, we assign each binary compound to one of the seven broad materials classes suggested by Figs.~\ref{fig: interpretation}a--g and evaluate $P(f \mid \text{class})$, the probability that a compound in a given class activates facet $f$. The resulting correlation matrix (Fig.~\ref{fig: interpretation}h) is strongly diagonal, indicating that each materials class is dominated by a single facet. This provides a compact quantitative summary of the materials organization captured by the max--affine representation and confirms that the facets of the convex hull correspond to physically meaningful thermodynamic regimes.

Within these regimes, the facet coefficients $a_{f,s}$ define the effective chemical potentials $\tilde\mu_s$ for all materials on facet $f$, so that members of a given facet share a common thermodynamic fingerprint. Materials on the same facet are therefore expected to exhibit similar trends in any property governed by energy differences. If the seven-facet structure reflects genuine thermodynamic families rather than merely an economical fit to formation energies, this organization should also manifest directly in observables such as defect energetics and elemental mixing behavior.

\subsection*{Bonding-geometry-free defect energies}

\begin{figure*}[htbp]
  \includegraphics[width=0.99\textwidth]{Figures_jpeg/defects.jpg}
    \caption{\textbf{DFT-calculated versus predicted defect energies.} 
    Perfect agreement, up to a constant offset, is indicated by the red dashed reference lines ($y=x+C$). These are true out-of-distribution predictions with no fitting other than a single constant shift. In all panels, max-affine values are shown as blue points computed using finite-difference approximations with step size $\delta=0.5$. 
    \textbf{a} DFT-computed interstitial, vacancy, and substitution energies (all defect classes) versus max-affine predictions (offset $C=$+1.24~eV/atom). 
    \textbf{b,c} DFT-computed interstitial energies versus predictions \eqref{eq: inf interstitial} for nitrogen (\textbf{b}) and oxygen (\textbf{c}) interstitials. Offsets of $C=+1.75$~eV/atom in \textbf{b} and $C=+1.24$~eV/atom in \textbf{c}. 
    \textbf{d} DFT-computed oxygen-vacancy energies versus predictions \eqref{eq: inf vacancy} for a range of metal oxides (offset $C=+1.28$~eV/atom). 
    \textbf{e} DFT-computed substitution energies versus predictions \eqref{eq: inf substitution} for a variety of impurities in Nb (offset $C=+0.22$~eV/atom).}

  \label{fig: defects}
\end{figure*}

If the facet-defined thermodynamic regimes reflect the underlying structure of the convex hull, they should also be reflected in defect energetics. We test this by  examining whether the max--affine representation---fitted only to bulk formation energies---reproduces systematic trends in DFT-calculated defect energies without refitting.  We consider substitutional, vacancy, and interstitial defects, which probe distinct lattice perturbations across diverse materials environments.

To evaluate defect energetics within our max--affine framework, we first define compositional chemical potentials using a notation in which species appear as subscripts and thermodynamic context as superscripts, $\tilde{\mu}_{\mathrm{[species]}}^{\mathrm{[context]}}$. We thus define the chemical potential of element $\mathrm{A}$ in the context of a material $\mathrm{M}$ as
\begin{equation} \label{eq:contextmu}
\tilde{\mu}_{\mathrm{A}}^{\mathrm{M}} \equiv \frac{\partial}{\partial x_A} \left( \max_f\left[\sum_s a_{f,s}\, x_s(\mathrm{M}) +b\right]\right) = a_{f_{\mathrm{M}},\mathrm{A}},
\end{equation}
where $\boldsymbol{x}({\mathrm{M}})$ denotes the composition of material $\mathrm{M}$ and $f_{\mathrm{M}}$ is the active facet for that composition. Here, $\mathrm{M}$ may represent a multicomponent material, for example $\mathrm{A}_x\mathrm{B}_{1-x}$. For a single-element material, $\mathrm{M} = \mathrm{A}$, we omit the superscript and write simply $\tilde{\mu}_{\mathrm{A}}$.

The energies of all three prototypical point defects—substitutional, vacancy, and interstitial—can be expressed entirely in terms of the effective compositional chemical potentials. Beginning with substitutional defects, the substitution energy quantifies the cost of replacing an atom in host material A with one of type B. In standard \emph{ab initio} calculations, the substitution energy is approximated using an $n$-atom periodic supercell and computed as
\begin{align}
    E_\mathrm{sub}^{n}(\mathrm{A}\leftarrow \mathrm{B}) = E(\mathrm{A}_{n-1} \mathrm{B}) + E(\mathrm{A})\notag \\ - (E(\mathrm{A}_n) + E(\mathrm{B})),
    \label{eq: n substitution}
\end{align}
where $E(\mathrm{M})$ represents the energy of one formula unit of M as written. In the dilute limit $n\to\infty$, corresponding to a single impurity in an infinite host, the substitution energy reduces to a compositional derivative of the formation energy,
\begin{equation}
    E_{\mathrm{sub}}^{\infty}(\mathrm{A}\leftarrow\mathrm{B}) = E(\mathrm{A}) - E(\mathrm{B}) + \partial_x^+ E(\mathrm{A}_{1-x}\mathrm{B}_x)|_{x=0},
    \label{eq: inf substitution}
\end{equation}
where $\partial_x^+$ denotes a forward derivative with respect to $x$ (see Methods for derivation). Within the max--affine framework, this derivative is determined directly by the facet-dependent chemical potentials. Substituting \eqref{eq:contextmu} yields the simple expression
\begin{equation}
    E_{\mathrm{sub}}^{\mathrm{MA}}(\mathrm{A}\leftarrow\mathrm{B})=\tilde{\mu}^{\mathrm{A}}_\mathrm{B}-\tilde{\mu}_{\mathrm{B}},
    \label{eq: MA-substitution}
\end{equation}
showing that the substitution energies are governed by differences of effective compositional chemical potentials.

Interstitial energies quantify the cost of inserting a foreign atom into an otherwise pristine lattice and are computed as
\begin{equation}
    E_{\mathrm{int}}^{n}(\mathrm{A}\leftarrow\mathrm{B}) = E(\mathrm{A}_n \mathrm{B}) - (E(\mathrm{A}_n) + E(\mathrm{B})).
    \label{eq: n interstitial}
\end{equation}
In the dilute limit $n\to\infty$ (see Methods), this reduces to
\begin{equation}
    E_{\mathrm{int}}^{\infty}(\mathrm{A}\leftarrow\mathrm{B}) = E(\mathrm{A}) - E(\mathrm{B}) + \partial_x^+ E(\mathrm{A}_{1-x}\mathrm{B}_x)|_{x=0},
    \label{eq: inf interstitial}
\end{equation}
which has the same mathematical form as the dilute-limit substitution energy, \eqref{eq: inf substitution}. As a result, within the max--affine framework, the interstitial energy is likewise given by
\begin{equation}
    E_{\mathrm{int}}^{\mathrm{MA}}(\mathrm{A}\leftarrow\mathrm{B}) = \tilde{\mu}_\mathrm{B}^{\mathrm{A}} - \tilde{\mu}_{\mathrm{B}}.
    \label{eq: MA interstitial}
\end{equation}

Vacancy energies quantify the cost of removing an atom of type A from an $\mathrm{A}_x \mathrm{B}_{1-x}$ host and are computed as
\begin{align}
    E_{\mathrm{vac}}^{n}(\mathrm{A}_x \mathrm{B}_{1-x}\rightarrow\mathrm{A})
    = E(\mathrm{A}) + E\bigl(\mathrm{A}_{x n-1}\mathrm{B}_{(1-x)n}\bigr)
     \notag \\ - E\bigl(\mathrm{A}_{x n}\mathrm{B}_{(1-x)n}\bigr).
    \label{eq: n vacancy}
\end{align}
In the dilute limit $n\to\infty$ (see Methods), this becomes
\begin{align}
    E_{\mathrm{vac}}^{\infty}(\mathrm{A}_x \mathrm{B}_{1-x}\rightarrow\mathrm{A})
    =& E(\mathrm{A}) - E(\mathrm{A}_x \mathrm{B}_{1-x}) \notag \\
    &-(1-x)\partial_x^-\,E\bigl(\mathrm{A}_x \mathrm{B}_{1-x}\bigr),
    \label{eq: inf vacancy}
\end{align}
where $\partial_x^-$ is a backward derivative. Substituting the max--affine model yields
\begin{align}
    E_{\mathrm{vac}}^{\mathrm{MA}}(\mathrm{A}_x &\mathrm{B}_{1-x}\rightarrow\mathrm{A})
    =\tilde{\mu}_{\mathrm{A}}-\tilde{\mu}^{\mathrm{A}_x\mathrm{B}_{1-x}}_\mathrm{B} \notag \\
    &-(1-2x)\left(\tilde{\mu}_{\mathrm{B}}^{\mathrm{A}_x\mathrm{B}_{1-x}} - \tilde{\mu}_{\mathrm{A}}^{\mathrm{A}_x\mathrm{B}_{1-x}}\right),
    \label{eq: inf MA vacancy}
\end{align}
showing that vacancy energies, like substitutional and interstitial defects, are likewise governed by differences of effective compositional chemical potentials.

To enable comparison with \emph{ab initio} density-functional theory (DFT), we evaluate Eqs.~(\ref{eq: inf substitution},\ref{eq: inf interstitial},\ref{eq: inf vacancy}) using the max--affine representation using finite-difference compositional derivatives (step size $\delta=0.5$). This partially mitigates the max--affine representation's bias toward phase-separated energetics. Because the parameters are determined exclusively from bulk formation energies, with no defect energies included, the results that follow are true out-of-distribution predictions.

To test whether this defect framework captures trends across distinct environments, we evaluate N and O interstitials in metallic hosts, substitutional defects in Nb, and oxygen vacancies in metal oxides. Across this compositionally diverse set, the max-affine framework reproduces the DFT trends (Fig.~\ref{fig: defects}a). Predicted defect energies exhibit systematic downward shifts, consistent with a near-hull bias inherited from training on bulk formation energies. These offsets are approximately constant within each defect class and are corrected by $+1.24$~eV/atom in the overall comparison (Fig.~\ref{fig: defects}a), $+1.75$~eV/atom for N interstitials (Fig.~\ref{fig: defects}b), $+1.24$~eV/atom for O interstitials (Fig.~\ref{fig: defects}c), $+1.28$~eV/atom for vacancies (Fig.~\ref{fig: defects}d), and $+0.22$~eV/atom for substitutions (Fig.~\ref{fig: defects}e). Although the absolute errors exceed DFT accuracy, the observed correlations are notable given the interpretability in terms of just seven context-dependent chemical potentials per element, the diversity of defect classes, and the fact that these are genuine out-of-distribution predictions.

These results show that the structure uncovered in the seven-facet convex hull extends beyond bulk stability to organize defect energetics across diverse materials systems. Although fitted solely to bulk formation energies, the representation captures systematic trends in defect energetics, indicating that they are governed by the same effective chemical potentials that define the facet structure.  The defect calculations therefore provide independent physical confirmation that a small number of thermodynamic families govern both bulk stability and defect energetics. From a practical perspective, this framework enables rapid screening of combinatorially large multicomponent compositional spaces for targeted defect properties without requiring prior identification of minimum-energy structures.

\subsection*{Elemental Spatial Correlations in High-Entropy Materials}
\begin{figure*}[htbp]
  \includegraphics[width=0.99\textwidth]{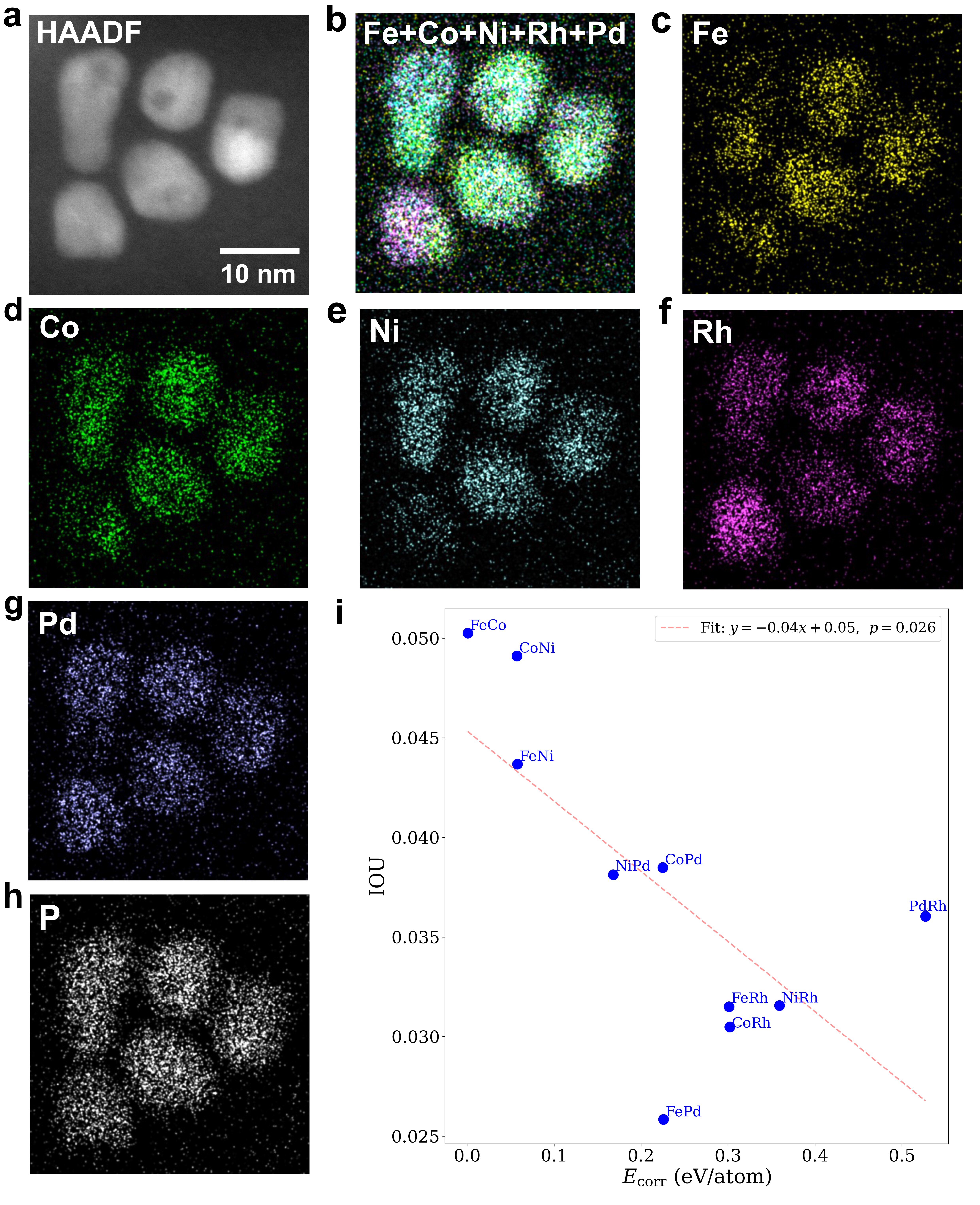}
\caption{\textbf{Elemental spatial correlations in high-entropy phosphide nanoparticles.} 
\textbf{a} High-angle annular dark-field (HAADF) STEM image of high-entropy phosphide nanoparticles. 
\textbf{b--h} Scanning transmission electron microscopy coupled with energy-dispersive X-ray spectroscopy (STEM–EDS) elemental maps of the $\mathrm{(FeCoNiRhPd)_2P}$ high-entropy phosphide, showing the spatial distribution of all metal elements in \textbf{b} and individual elements in \textbf{c--h}, respectively, for the same nanoparticles shown in \textbf{a}. All elemental maps are acquired from the same particles and are spatially registered to the HAADF image. 
\textbf{i} Intersection-over-union (IOU) values (\eqref{eq: iou}, threshold 0.55) extracted from the STEM–EDS images, plotted against the correlation energy $E_{\mathrm{corr}}$ descriptor \eqref{eq: correlation} computed using the max-affine representation. The red dashed line denotes the line of best fit which exhibits a p-value of 0.026, indicating statistical significance.}

  \label{fig: correlations}
\end{figure*}

We next ask whether the seven-family organization extends to elemental spatial correlations in compositionally complex high-entropy materials. If this structure is indeed general, it should also govern how elements preferentially mix or segregate in multicomponent environments. Figures~\ref{fig: correlations}a--h show HAADF-STEM and STEM--EDS data for $(\mathrm{FeCoNiRhPd})_2\mathrm{P}$ high-entropy phosphide nanoparticles. The elemental maps reveal clear deviations from random mixing, including strong correlation between Co- and Ni-rich regions (Figs.~\ref{fig: correlations}d,e) and strong anticorrelation between Fe- and Rh-rich regions (Figs.~\ref{fig: correlations}c,f), along with additional nonrandom patterns across the nanoparticles.

To quantify these correlations, we represent each elemental map as $I_\alpha(p)$, where $p$ labels pixels and $\alpha$ denotes an element. After normalizing intensities to $[0,1]$ and thresholding at 0.55, we compute the intersection--over--union \cite{jaccard1901}
\begin{equation}
    \mathrm{IOU}(\alpha,\beta)=\frac{I_\alpha\cap I_\beta}{I_\alpha\cup I_\beta},
    \label{eq: iou}
\end{equation}
which measures the spatial overlap between elements $\alpha$ and $\beta$. High IOU indicates correlation, while low IOU indicates anticorrelation.

Such spatial correlations reflect local energetic competition between elements for a shared bonding environment. This competition is directly encoded in composition-only formation energies, which can be used to probe a broad range of material behaviors through appropriate energy differences. To characterize correlations between two elements A and B in a high-entropy phosphide, we consider the formation energies of ternary compositions $\mathrm{A}_{2x/3}\mathrm{B}_{2(1-x)/3}\mathrm{P}_{1/3}$. Empirically, the representation exhibits a nearly linear dependence on $x$, indicating simple enthalpic mixing without energetically preferred intermediate compositions. In this regime, the slope of the energy with respect to composition determines mixing behavior: a large absolute slope reflects a strong enthalpic preference for phase separation into binary phosphides, while a small slope indicates weak enthalpic driving forces, allowing entropy to stabilize mixed configurations exhibiting spatial correlation. This leads to a simple descriptor for the degree of correlation between species pairs,
\begin{equation}
    E_{\mathrm{corr}}(\mathrm{A}, \mathrm{B}) =
    \bigl| E(\mathrm{A}_{2/3}\mathrm{P}_{1/3}) -
           E(\mathrm{B}_{2/3}\mathrm{P}_{1/3}) \bigr|,
    \label{eq: correlation}
\end{equation}
which measures the magnitude of the compositional energy slope within the max–affine framework and thus the tendency for separation and spatial anticorrelation.

Figure~\ref{fig: correlations}i compares $E_{\mathrm{corr}}$ from the max--affine representation with the experimental IOU values. We find that larger $E_{\mathrm{corr}}$ indeed corresponds to spatial anticorrelation and thus lower IOU, with a statistically significant correlation ($p=0.026$). Pairs such as Co--Ni, Fe--Co, and Fe--Ni are correctly predicted by the max-affine framework to be correlated, whereas Fe--Rh, Ni--Rh, and Co--Rh are correctly predicted to be anticorrelated. 

No parameters have been fitted to the experimental data, making this a genuine out-of-sample test. These results show that the same low-dimensional thermodynamic structure that organizes the convex hull also captures the dominant energetic drivers of elemental mixing and segregation in high-entropy materials. The seven-facet structure is therefore not limited to bulk formation and point-defect energies, but extends to emergent spatial correlations in compositionally complex materials.

\section*{Discussion}
In this work, we have shown that the thermodynamic stability of inorganic materials across composition space exhibits a striking and previously unrecognized simplicity. The convex hull of formation energies exhibits a low-faceted structure that is well described, at near DFT accuracy, by a max–affine representation with only seven linear facets, corresponding to seven coefficients per element. The result is a highly interpretable framework whose parameters are simply a small set of material-context dependent effective chemical potentials for each element. Importantly, the rapid saturation of approximation error with facet number suggests that these seven facets are not merely an efficient parameterization, but instead reflect a small set of underlying, dominant materials families that govern thermodynamic stability across multi-element materials classes. Finally, the coefficients defining each facet have a natural interpretation as distinct sets of effective compositional chemical potentials, ensuring consistent materials behavior aligned with periodic-table trends for the materials on each facet.

Beyond reproducing bulk thermodynamic stability, the same seven-facet structure is reflected across distinct physical observables. Without retraining or structural input, this representation captures systematic trends in defect formation energies and reproduces experimentally observed elemental correlations in high-entropy materials. These results indicate that the underlying thermodynamic structure is not specific to formation energies, but reflects a shared set of effective chemical potentials that govern energy differences across bulk stability, defect energetics, and elemental mixing. The agreement across these qualitatively different settings therefore provides independent physical evidence that a small number of thermodynamic families organizes a broad range of material properties.

At the same time, the simplicity and transferability of this representation make it practically useful for navigating large composition spaces. Because it depends only on composition and involves a small number of interpretable parameters, it can be evaluated rapidly across vast combinatorial spaces without the need for explicit structural information. While it does not provide quantitatively precise defect energies, it enables efficient identification of material classes with favorable combinations of properties---such as low-energy defects, desired substitutional tendencies, or correlated elemental mixing---thereby serving as a first-pass screening tool that can guide more detailed structural calculations. In this way, the low-dimensional thermodynamic structure revealed here not only organizes materials understanding, but also provides a practical route to accelerated materials exploration.

Despite its broad applicability, the present formulation is not without limitations. Because the representation is determined primarily by near-hull data, it introduces systematic stabilization biases, necessitating finite-difference derivatives and constant offsets when comparing absolute defect energies to DFT. Sparse sampling in certain regions of composition space can blur facet assignments, and explicit finite-temperature entropies are not included. These limitations, however, suggest clear and tractable extensions, such as calibrating offsets using small reference datasets or attaching uncertainty estimates to identify out-of-distribution compositions and guide targeted data acquisition.

By exposing the geometric structure underlying the formation-energy landscape, the max–affine framework reveals how inorganic thermodynamics is organized into a small number of meaningful families rather than an arbitrarily complex continuum. This organization is directly reflected in the 92-dimensional formation-energy landscape, which is well approximated by a low-faceted polyhedral surface. Viewed in this light, the long-standing intuition that materials fall into distinct families with shared behavior is not merely a useful classification, but a reflection of the underlying geometry of the convex hull of formation energies.

\section{Methods}
\subsection*{Dataset Preparation}
To construct the dataset used in the max-affine construction, we start from bulk inorganic compounds in the Materials Project database~\cite{materials_project} with density functional theory (DFT)--computed formation energies. For each composition, only the lowest-energy structure is retained. We then restrict the dataset to materials containing between one and five elements and discard any compound lying more than 0.025~eV/atom above the convex hull, yielding a final dataset of 51{,}068 compounds. This dataset is randomly split into 80\% fitting and 20\% testing sets.

\subsection*{Max--affine construction}
The parameters of the max--affine representation are determined by minimizing the mean squared error using the Adam optimizer~\cite{adam} (learning rate $1\times10^{-3}$) implemented in PyTorch~\cite{pytorch}, with a batch size of 512. The optimization is run for 1{,}600 epochs to convergence.  In practice, the representation is constructed using a softened version of \eqref{eq: max-affine},
\begin{equation}
    E(\boldsymbol{x})=\tau \log\left(\sum_{f=1}^F \exp\left[\frac{\sum_{s=1}^{92}a_{fs}x_s}{\tau}\right]\right) + b,
    \label{eq: smooth-max-affine}
\end{equation}
where $\tau>0$ is a temperature parameter. During optimization, $\tau$ is gradually decreased to approach the hard max--affine form of \eqref{eq: max-affine}. The $\tau \rightarrow 0$ limit, when \eqref{eq: smooth-max-affine} becomes \eqref{eq: max-affine}, is taken after optimization.

\subsection*{Ab initio Calculations}

All density functional theory (DFT) calculations of point defects were carried out using the \texttt{JDFTx}~\cite{jdftx} code. For interstitial and vacancy formation energies, $2\times2\times2$ or $3\times3\times3$ supercells (depending on the parent lattice symmetry) were generated from conventional unit cells obtained via the Materials Project API. Substitution energies were evaluated in Nb supercells with a single metal replacement, and oxygen vacancies were modeled by removing one O atom from a $2\times2\times2$ metal oxide supercell. Ionic cores were described using the GBRV ultrasoft PBE potentials for interstitial and vacancy calculations and the SG15 ONCV PBE library for substitution energies~\cite{PBE_functional,schlipf2015}. Plane-wave cutoffs of 20~H for wavefunctions (100~H for charge density) were used for interstitial and vacancy calculations, while cutoffs of 30~H (200~H for charge density) were used for substitution energies, ensuring total-energy convergence to within $10^{-5}$~H. A $\boldsymbol{k}$-point grid of $4\times4\times4$ was employed, together with a Fermi--Dirac smearing corresponding to an effective temperature of $4\times10^{-3}$~H. Complete information about all calculations—including final atomic coordinates, total energies, and input parameters—is available on the GitHub repository.

\subsection*{Defect Energies from Composition-Only Formation Energies}

We first derive the substitution energy expression in \eqref{eq: inf substitution}. Starting from the finite-$n$ definition in \eqref{eq: n substitution} and using the extensivity of the energy,
$E(xM)=x\,E(M)$, we rewrite the substitution energy as
\begin{align}
    E_{\mathrm{sub}}^{n}(\mathrm{A}\leftarrow \mathrm{B}) &=E(\mathrm{A}_{n-1} \mathrm{B}) + E(\mathrm{A})- (E(\mathrm{A}_n) + E(\mathrm{B})) \notag\\
    &= E(\mathrm{A}) - E(\mathrm{B}) \notag \\
    &\quad + n\!\left[E\!\left(\mathrm{A}_{1-\frac{1}{n}}\mathrm{B}_{\frac{1}{n}}\right) - E(\mathrm{A})\right].  \label{eq:subasympt}
\end{align}
Introducing $\delta = 1/n$, we expand the energy about $\delta=0$ and take the $n\rightarrow\infty$ (i.e., $\delta\rightarrow 0^+$) limit,
\begin{align*}
    E_{\mathrm{sub}}^{\infty}(\mathrm{A}\leftarrow\mathrm{B})
    &= E(\mathrm{A}) - E(\mathrm{B}) \notag \\
    &\quad + \frac{1}{\delta}
    \Big[
        E(\mathrm{A}_{1-\delta}\mathrm{B}_\delta) - E(\mathrm{A})
    \Big]\notag\\&\quad\quad + \mathcal{O}(\delta). 
\end{align*}
Taking the limit $\delta\rightarrow 0^+$ yields
\begin{equation*}
    E_{\mathrm{sub}}^{\infty}(\mathrm{A}\leftarrow\mathrm{B})
    = E(\mathrm{A}) - E(\mathrm{B})
    + \partial_\delta^+ E(\mathrm{A}_{1-\delta}\mathrm{B}_\delta)\big|_{\delta=0},
\end{equation*}
recovering \eqref{eq: inf substitution}. We note that a \emph{forward} derivative $\partial^+$ must be used, since the perturbation $\delta=1/n>0$. 

We next derive the dilute-limit interstitial energy expression in \eqref{eq: inf interstitial}. Starting from a bulk host containing $n-1$ atoms and again using extensivity, the finite-size interstitial energy \eqref{eq: n interstitial} is
\begin{align*}
    E_{\mathrm{int}}^{n-1}(\mathrm{A}\leftarrow\mathrm{B})
    &= E(\mathrm{A}_{n-1}\mathrm{B}) - \bigl(E(\mathrm{A}_{n-1}) + E(\mathrm{B})\bigr) \\
    &=E(\mathrm{A}_{n-1} \mathrm{B}) + E(\mathrm{A})- (E(\mathrm{A}_n) + E(\mathrm{B})),
\end{align*}
which is identical in form to the starting point of \eqref{eq:subasympt}. This explains why the substitutional and interstitial energies in the main text, \eqref{eq: inf substitution} and \eqref{eq: inf interstitial}, have exactly the same form in the dilute limit $n\rightarrow\infty$. Physically, the reason why the same mathematical form arises is that the ground-state formation energy function $E(\boldsymbol{x})$ implicitly encodes the lowest-energy mechanism by which an impurity is incorporated into the host lattice, whether substitutional or interstitial.

Finally, we derive the vacancy--energy expression in \eqref{eq: inf vacancy}. 
Starting from the finite-$n$ form in \eqref{eq: n vacancy}, we rewrite it using extensivity as
\begin{align*}
    E_{\mathrm{vac}}^{n}&(\mathrm{A}_x \mathrm{B}_{1-x}\rightarrow\mathrm{A})\\
    &= E(\mathrm{A}) + E\bigl(\mathrm{A}_{x n-1}\mathrm{B}_{(1-x)n}\bigr)
     - E\bigl(\mathrm{A}_{x n}\mathrm{B}_{(1-x)n}\bigr), \\&=(n-1)E\left(\mathrm{A}_{x - \frac{1-x}{n-1}}\mathrm{B}_{1-x + \frac{1-x}{n-1}}\right) \\
     &\qquad\qquad\qquad\qquad+ E(\mathrm{A}) - nE(\mathrm{A}_{x}\mathrm{B}_{1-x}).
\end{align*}
Note that $(1-x)/(n-1)\ge 0$, so the physically admissible perturbation always \emph{decreases} the fraction of A and \emph{increases} the fraction of B. 

Defining $\delta = 1/(n-1)>0$ and expanding about $\delta=0$ yields
\begin{align*}
    E_{\mathrm{vac}}^{\infty}(\mathrm{A}_x \mathrm{B}_{1-x}\rightarrow\mathrm{A})
    &= E(\mathrm{A}) - E(\mathrm{A}_x\mathrm{B}_{1-x}) \notag \\
    &\quad
    - (1-x)\,\partial_x^- E(\mathrm{A}_x\mathrm{B}_{1-x})
    + \mathcal{O}(\delta),
\end{align*}
where $\partial_x^-$ denotes the \emph{backward} derivative with respect to $x$. Here, the backward derivative is required because the admissible perturbation moves leftward along the composition axis---
a forward or central difference would probe compositions to the right of $x$, which are inaccessible under this construction and would therefore misrepresent the relevant local slope of the formation--energy curve. Finally, discarding $\mathcal{O}(\delta)$ terms recovers \eqref{eq: inf vacancy}.

\subsection*{Synthesis of high-entropy phosphide nanoparticles}
First, precursor solutions were prepared by dissolving each metal salt [36.5 mg Co(acac)$_3$, 27.0 mg Ni(acac)$_2$, 42.2 mg Rh(acac)$_3$, 32.2 mg PdCl$_2$, 36.5 mg Fe(acac)$_3$] in 10 mL of oleylamine with gentle heating and sonication. Next, a 50-mL three-neck round-bottom flask was assembled with a thermocouple, reflux condenser, gas flow adapter, and rubber septum; 10 mL of 1-octadecene (ODE) and 6 mL of oleylamine (OLAM) were added and degassed under vacuum at 110 $^\circ$C for 1 hour. The system was then purged by alternating argon and vacuum three times, followed by heating under a continuous argon flow to 320 $^\circ$C. After this step, 1 mL of the Ni(acac)$_2$ solution and 0.75 mL of each of the remaining metal precursor solutions were combined, degassed under vacuum for 15 minutes, purged three times with argon and vacuum, and rapidly injected into the 50 mL reaction flask containing ODE and OLAM at 320 $^\circ$C. The reaction was held at 315 $^\circ$C for 1.5 hours, cooled to 220 $^\circ$C, and quenched in a room-temperature water bath. The product was isolated by washing twice with a 1:2 (v/v) mixture of toluene and ethanol, followed by centrifugation and redispersion in toluene after each wash. The purified nanoparticles were suspended and stored in toluene. High-angle annular dark-field scanning transmission electron microscopy (HAADF-STEM) images and corresponding STEM energy-dispersive X-ray spectroscopy (STEM–EDS) elemental maps were acquired using an FEI Talos F200X S/TEM operated at an accelerating voltage of 200 kV. The STEM–EDS data were processed and analyzed using Velox 3.6.0 software. Elemental mapping and analysis were performed using the Fe K$\alpha$, Co K$\alpha$, Ni K$\alpha$, Rh L$\alpha$, and Pd L$\alpha$ characteristic energy lines.

\section{Data Availability}
The datasets generated and/or analysed during the current study are available in the Max-Affine-Data repository, https://github.com/jt577/Max-Affine-Data. 

\section{Code Availability}
The underlying code for this study is available in Max-Affine-Data and can be accessed via this link https://github.com/jt577/Max-Affine-Data.

\bibliographystyle{unsrtnat}
\bibliography{refs}

\section{Acknowledgements}
J.T., S.K., R.S., H.A., and T.A. disclose support for the research of this work from the Center for Alkaline-Based Energy Solutions (CABES), part of the Energy Frontier Research Center (EFRC) program supported by the U.S. Department of Energy under grant DE-SC-0019445. S.C. and C.M. disclose support for the research of this work from the US National Science Foundation (NSF) under award PHY-1549132, the Center for Bright Beams (CBB).

\section{Author Contributions}
J.T. conceived and fit all representations, and edited the manuscript. S.C. derived the vacancy energy expression and helped J.T. in predicting vacancy energies. C.M. and R.C. performed defect DFT calculations. S.K. and R.S. synthesized and imaged the high entropy phosphide nanoparticles, and edited the corresponding sections of the manuscript. H.A. and P.F. supervised the study. T.A. conceived and supervised the study, assisted in interpreting the results, and edited the manuscript. All authors reviewed the manuscript.

\section{Competing Interests}
All authors declare no financial or non-financial competing interests.

\end{document}